\documentclass[preprint,aps,prc,tightenlines,floatfix,showpacs,showkeys,nofootinbib]{revtex4}

\usepackage{graphicx}
\usepackage{dcolumn}
\usepackage{bm}

\def\sss{\mbox{\boldmath $\sigma$}}

\newcommand{\be}{\begin{equation}}
\newcommand{\ee}{\end{equation}}
\newcommand{\bea}{\begin{eqnarray}}
\newcommand{\eea}{\end{eqnarray}}

\newcommand{ \bb }{$2\nu\beta\beta$}
\newcommand{ \bbm }{$2\nu\beta^-\beta^-$}
\newcommand{ \bbO }{$0\nu\beta\beta$}

\begin{document}

\topmargin -0.50in

\title{ Origin of 
a sensitive dependence of calculated $\beta\beta$-decay amplitudes on 
particle-particle residual interaction}

\author{Vadim Rodin}
\email{vadim.rodin@uni-tuebingen.de}
\affiliation{Institut f\"{u}r Theoretische Physik der Universit\"{a}t
T\"{u}bingen, D-72076 T\"{u}bingen, Germany}
\author{Amand Faessler}
\affiliation{Institut f\"{u}r Theoretische Physik der Universit\"{a}t
T\"{u}bingen, D-72076 T\"{u}bingen, Germany}

\begin{abstract}

In the present work the sensitivity of 
calculated $\beta\beta$-decay amplitudes to a realistic residual interaction is analyzed in the framework of the approach of Refs.~\cite{Rum98,Rodin05}. Both the Gamow-Teller (GT) and Fermi (F) matrix elements $M^{2\nu}$ for two-neutrino $\beta\beta$ decay (\bb decay), along with the monopole transition contributions to the total matrix elements $M^{0\nu}$ of neutrinoless $\beta\beta$ decay (\bbO decay), are calculated within the quasiparticle random-phase approximation (QRPA). 
Decompositions of $M^{2\nu}$ and $M^{0\nu}$ are obtained by the method of Refs.~\cite{Rum98,Rodin05} in terms of the corresponding energy-weighted sum rules $S$. It is shown that in most of the cases almost the whole dependence of $M^{2\nu}$ and $M^{0\nu}$ on the particle-particle (p-p) renormalization parameter $g_{pp}$ is accounted for by the $g_{pp}$ dependence of the corresponding sum rules $S$.
General expressions relating $S$ to a realistic residual particle-particle interaction are derived, which show a pronounced sensitivity of $S$ to the singlet-channel interaction in the case of F transitions, and to the triplet-channel interaction in the case of GT transitions. 
Thus, the sensitivity of $M^{2\nu}$ and $M^{0\nu}$ to the SU(4)-symmetry-breaking part of the p-p residual interaction is dictated by the generic structure of the $\beta\beta$-decay amplitudes. Therefore, a choice of this part in a particular calculation needs a special caution.
Finally, a better isospin-consistent way of renormalization of a realistic residual p-p interaction to use in QRPA calculations is suggested.

\end{abstract}



\date{\today}
\maketitle
\section{Introduction}

Study of neutrinoless double beta ($0\nu\beta\beta$) decay provides one of a few ways to probe the absolute neutrino mass scale with a high sensitivity~\cite{vogelbook,fae98}.
To be able to deduce the effective Majorana neutrino mass from the measured half-lives of the decay, reliably calculated nuclear matrix elements (NME) $M^{0\nu}$ are needed in addition. 

The NME $M^{0\nu}$ have been calculated in different approaches: the quasiparticle random-phase approximation (QRPA)~\cite{Rod05,anatomy,src09} (including the recent version of the QRPA accounting for deformation~\cite{Fang10}),
the nuclear shell model (SM)~\cite{Poves}, the projected Hartree-Fock-Bogoliubov method (PHFB)~\cite{Rath}, the interacting boson model (IBM-2)~\cite{Iachello}, and the generator coordinate method with particle number and angular momentum projection (GCM+PNAMP)~\cite{Rodri10}. There has been great progress in the calculations over the last decade, and now $M^{0\nu}$ of different groups (apart from the SM one) seem to converge. However, $M^{0\nu}$ of the SM are systematically and substantially (up to a factor of 2) smaller than the results of the other approaches. This discrepancy calls for a better understanding of the aspects of nuclear structure that affect largely the calculated NME.

All the QRPA calculations of the nuclear $\beta\beta$-decay amplitudes have revealed 
their sensitivity to the strength of the particle-particle (p-p) interaction in the triplet channel,\footnote{The so-called ``$g_{pp}$ problem", whereby the factor $g_{pp}$ is used to renormalize the p-p interaction in the QRPA equations} which is mainly due to a high $g_{pp}$ sensitivity of the contribution of transitions through the intermediate $1^+$ states. 
Such a behavior of the two-neutrino $2\nu\beta\beta$-decay NME $M^{2\nu}$ was found for the first time in 
Ref.~\cite{Vog86}.
Reference~\cite{Vog86} contains a clue that the sensitivity may be related to the restoration of the Wigner spin-isospin SU(4) symmetry in nuclei.
The authors of Ref.~\cite{Vog86} also showed that the sensitivity
is not an artifact of the QRPA but also shows up in an exactly soluble schematic model.

An idea to use the concept of softly broken SU(4) symmetry 
as a basis for describing \bb-decay amplitude was put forward in Ref.~\cite{Rum98}. Because $M^{2\nu}$ vanishes in the limit when the SU(4) symmetry is exact, it is natural to express $M^{2\nu}$ explicitly in terms of those parts of the nuclear Hamiltonian $\hat H$ that are responsible for the violation of the symmetry. An identity transformation introduced in Ref.~\cite{Rum98} allows one to shed light on the general properties of the \bb-decay amplitude.
However, the computational realization of this idea in Ref.~\cite{Rum98} made use of a oversimplified model of independent quasiparticles and, therefore, could not address the question of the $g_{pp}$ sensitivity.

The next step was made in Ref.~\cite{Rodin05}, where the basic concept of Ref.~\cite{Rum98} was applied in the framework of a QRPA model.
As in Ref.~\cite{Rum98}, the starting point for the analysis was a model-independent, identity, transformation of $M^{2\nu}$. 
That allowed the authors to partition $M^{2\nu}$ into two terms which are sensitive to different parts of $\hat H$. The dominating source of the $g_{pp}$ sensitivity was associated in Ref.~\cite{Rodin05} with a specific energy-weighted sum rule $S$ for double beta decay, that depends exclusively on the residual p-p interaction. An analytical representation for $S$ was obtained in Ref.~\cite{Rodin05} within the QRPA for the case of a simple separable p-p interaction, which showed that $S$ vanishes 
at the point where the SU(4) symmetry is restored in the p-p sector of the model Hamiltonian. The rest of $M^{2\nu}$ was shown to behave smoother on $g_{pp}$ for the realistic values of the p-p interaction strength, and was mainly determined by the other important source of breaking the SU(4) symmetry, namely the spin-orbit part of the nuclear mean field.

In the present work we apply the approach of Refs.~\cite{Rum98,Rodin05} to analyze 
the sensitivity of calculated $\beta\beta$-decay amplitudes to a realistic residual interaction. Both the Gamow-Teller (GT) and Fermi (F) matrix elements $M^{2\nu}$, along with the monopole transition contributions to $M^{0\nu}$, are calculated within the QRPA, making use of a realistic residual interaction (the Brueckner $G$ matrix) as described in Refs.~\cite{Rod05,anatomy,src09}.
General expressions relating the sum rule $S$ to a realistic residual p-p interaction are derived, which show a pronounced sensitivity of $S$ to the singlet-channel of a two-body interaction in the case of F transitions, and to the triplet-channel of a two-body interaction in the case of GT transitions. 
In this connection, $S$ for GT transitions would be a better quantity for fitting $g_{pp}$ than the experimental $M^{2\nu}$, provided $S$ could be measured.~\footnote{Realistically, the absolute value of $S$ can only be determined experimentally (from charge-exchange reactions or single-$\beta$ decays) if the single-state dominance is realized in \bb-decay of one or another nuclear system.}. 
Identity partitions of $M^{2\nu}$, as well as the monopole transition contributions to $M^{0\nu}$, are obtained by the method of Refs.~\cite{Rum98,Rodin05}. It is shown that in most of the cases almost the whole $g_{pp}$ dependence of $M^{2\nu}$ and $M^{0\nu}$ can be attributed to the $g_{pp}$ dependence of the corresponding sum rules $S$. Thus, the sensitivity of $M^{2\nu}$ and $M^{0\nu}$ to the SU(4)-symmetry-breaking part of the p-p residual interaction is unavoidable since it is simply dictated by the generic structure of the $\beta\beta$ amplitudes. Finally, a better isospin-consistent way of a renormalization of a realistic residual p-p interaction to use in QRPA calculations is suggested.

\section{Identity partition of $\beta\beta$-decay amplitudes}

We adopt here the same line of reasoning as presented in Refs.~\cite{Rum98,Rodin05}.
The F and GT transitions are treated in a uniform way, and the single-particle (s.p.) operator $\hat \beta^{\pm}_J=\sum_a g_J(a)\tau^{\pm}(a)$ governs allowed Fermi
($g_{J=0}=1$) or GT ($g_{J=1}=\sss$) $\beta$--transitions, respectively. Here, $J=0,1$ is the angular momentum of a state (with positive parity) that can be connected by the operator $\hat \beta^{\pm}_J$ with the ground state of an even-even nucleus.

The {\bbm}-decay amplitude can be written in the form 
~\cite{vogelbook,fae98}
\be
\label{Mbb}
M^{2\nu}_J = \sum_s \frac{g_s f_s}{\bar\omega_s}\ .
\ee
Here,  $g_s=\langle 0^+_f \| \hat \beta^{-}_J\|J^+, s\rangle$ and 
$f_s=\langle J^+,s\|\hat \beta^{-}_J \| 0^+_i \rangle$ are the one-leg transition matrix elements of the operator $\hat \beta^{-}$
between an intermediate state $s$ of the isobaric nucleus $(N-1,Z+1)$
and the ground states (g.s.) of the parent ($Z,N$) and the final ($Z+2,N-2$) nucleus, respectively.
The energy denominator $\bar\omega_s$ in Eq.~(\ref{Mbb}) is the excitation energy 
of the $s$th state relative to the mean g.s. energy of the initial and final nucleus, $\bar\omega_s=(\omega_{s(i)}+{\omega}_{s(f)})/2$, with $\omega_{s(i)}=E_s-E_{0_i}$ ($\omega_{s(f)}=E_s-E_{0_f}$) representing the excitation energy of the $s$'th state relative to the g.s. of the initial (final) nucleus.

The following 
partition of $M^{2\nu}$ (\ref{Mbb}) can be performed~\cite{Rodin05}:
\bea
& M^{2\nu}_J={M'}^{2\nu}_J+\displaystyle\frac{S^{2\nu}_J}{\bar\omega_g^2} \label{Mbb1}\\
& {M'}^{2\nu}_J = \displaystyle\sum_s \frac{(\bar\omega_g^2-\bar\omega_s^2)g_s f_s}{\bar\omega_g^2\bar\omega_s} \label{Mbb'}\\
& S^{2\nu}_J=\displaystyle\sum_s \bar\omega_s g_s f_s, \label{S2nu}
\eea
where $\bar\omega_g$ is, in principle, an arbitrary energy. Note, that the second term in the r.h.s. of Eq.~(\ref{Mbb1}) is proportional to $S^{2\nu}_J$, which has the form (\ref{S2nu}) of an energy-weighted sum rule. One can get the following expression for $S^{2\nu}_J$~\cite{Rodin05}:
\be
S^{2\nu}_J =-\frac{1}{2} \langle 0^+_f| \hat S^{--}_J| 0^+_i\rangle \ ,\ \ \ \hat S^{--}_J\equiv\left[\widetilde {\hat \beta^{-}_J},\left[\hat \beta^{-}_J,\hat H\right]\right],
\label{S}
\ee
where $\hat H$ is the nuclear Hamiltonian and the tilde denotes the time-reversal operation. A sum over all the spin components is assumed in the case of the GT transitions. The expression (\ref{S}) 
contains explicit information about symmetry properties of the nuclear Hamiltonian in terms of the corresponding commutators.
If one chooses $\bar\omega_g$ in Eq.~(\ref{Mbb1}) to coincide with the energy $\bar\omega_G$ of the corresponding giant resonance (the isobaric analog state (IAS) or the Gamow-Teller resonance (GTR)), one sees immediately that both ${M'}^{2\nu}_J$ and $S^{2\nu}_J$ vanish in the limit of the exact isospin SU(2) ($J=0$) or spin-isospin SU(4) 
($J=1$) symmetries. 
It can be seen for small deviations from a symmetry that $S^{2\nu}_J$ depends linearly on the symmetry-breaking terms of the Hamiltonian, whereas ${M'}^{2\nu}_J$ has a weaker, quadratic, dependence on them.

It is shown in Ref.~\cite{Rodin05} that ${M'}^{2\nu}_J$ and $S^{2\nu}_J$ are sensitive to different terms in $\hat H$. By making use of the quasiboson approximation (QBA), $S^{2\nu}_J$ was demonstrated to be only determined by the p-p part of the residual interaction (chosen in that work in a separable form). 
An analytical expression for the GT sum rule $S^{2\nu}_{1}$ was also derived in Ref.~\cite{Rodin05}:
\be
S^{2\nu}_{1}=3 \displaystyle\frac{\Delta_n\Delta_p}{G_{0}}(1-g'_{pp}), 
\label{S_separ}
\ee
where $\Delta_n$ and $\Delta_p$ are the pairing gaps for neutrons and protons, and $g'_{pp}=\frac{G_{1}}{\bar G_{0}}$ is a ratio of the strength of the triplet (spin $S=1$) p-p interaction $G_1$ and the singlet (spin $S=0$) one $G_0$ (the latter governs the pairing correlations in nuclei).
In the derivation of Eq.~(\ref{S_separ}) the BCS vacua were taken the same for initial and final nuclei.
The point $g'_{pp}=1$, where $S^{2\nu}_{1}=0$, corresponds to the restoration of 
the SU(4) symmetry in the p-p sector of the model Hamiltonian of Ref.~\cite{Rodin05}.

It is useful to consider also a closure {\bb}-decay matrix element $M^{2\nu}_{J~\{cl\}}$ 
\be
\label{Mbb_cl}
M^{2\nu}_{J~\{cl\}} = \sum_s g_s f_s 
\ee
and to apply to it a partition similar to Eq.~(\ref{Mbb1}):
\bea
& M^{2\nu}_{J~\{cl\}}={M'}^{2\nu}_{J~\{cl\}}+\displaystyle\frac{S^{2\nu}_J}{\bar\omega_g} \label{Mbb1_cl}\\
& {M'}^{2\nu}_{J~\{cl\}} = \displaystyle\sum_s \frac{(\bar\omega_g-\bar\omega_s)g_s f_s}{\bar\omega_g}. \label{Mbb'_cl}
\eea

The nuclear matrix element $M^{0\nu}$ of \bbO decay is given by a sum of the partial amplitudes $M^{0\nu}_{s}(J)$ of transitions via the intermediate states of all multipolarities $J^\pi$ (see, e.g.,  Refs.~\cite{Rod05,anatomy,src09}):
\be
M^{0\nu} =\sum_{J} M^{0\nu}(J);\ M^{0\nu}(J)=\sum_{s}M^{0\nu}_{s}(J)\label{M--}
\ee
$M^{0\nu}_{J}$ also can be partitioned:
\bea
&M^{0\nu}(J)={M'}^{0\nu}(J)+\displaystyle\frac{S^{0\nu}_J}{\bar\omega_g}\label{M0nu}\\
&{M'}^{0\nu}(J)= \displaystyle\sum_s \frac{(\bar\omega_g-\bar\omega_s) M^{0\nu}_{s}(J)}{\bar\omega_g};\ \ 
S^{0\nu}_J=\displaystyle\sum_s \bar\omega_s M^{0\nu}_{s}(J)\label{M0nu1}
\label{S0nu}
\eea
Below only the contributions $M^{0\nu}(0^+),\ M^{0\nu}(1^+)$ of the monopole transitions through the intermediate states $0^+,1^+$ are analyzed, which are known to be sensitive to the p-p interaction.

We stress again that the all the transformations of 
the $\beta\beta$-decay amplitudes Eqs.(\ref{Mbb1})--(\ref{S0nu}) introduced above are identical, and therefore do not rely on any nuclear model.

\section{Derivation of $S^{2\nu}_J$ for a realistic residual interaction}

In this section we present a derivation of $\hat S^{--}_J$ and $S^{2\nu}_J$~(\ref{S}) 
in the case of a general realistic residual interaction.

One sees immediately that a single-particle mean field, as containing only isoscalar and isovector terms, exactly drops out of the double commutator~(\ref{S}) defining $\hat S^{--}_J$. 
Only the residual two-body interaction 
$\hat V= \frac12 \displaystyle\sum_{a\neq b} v_{ab}$ 
contributes (we work here in the first quantization since it simplifies the further derivation):
\bea
\ \hskip -1cm\hat S^{--}_J&=&\frac12 \sum_{a\neq b} S^{2\nu}_J(ab)\\
\ \hskip -1cm S^{2\nu}_J(ab) &\equiv& \left[{\tilde\beta}^{-}_{ab,J},\left[\beta^{-}_{ab,J},v_{ab}\right]\right]=
{\tilde\beta}^{-}_{ab,J}\beta^{-}_{ab,J} v_{ab} + v_{ab}\beta^{-}_{ab,J}{\tilde\beta}^{-}_{ab,J} -
\beta^{-}_{ab,J} v_{ab} {\tilde\beta}^{-}_{ab,J}-{\tilde\beta}^{-}_{ab,J} v_{ab} \beta^{-}_{ab,J}.\label{S_J(ab)}
\eea
Here, $\beta^{\pm}_{ab,J} \equiv g_J(a)\tau^{\pm}(a)+g_J(b)\tau^{\pm}(b)=
g_J^+ T^{\pm}+g_J^- t^{\pm}$, $g_J^+ = \frac12(g_J(a)+g_J(b))$, $g_J^- = \frac12(g_J(a)-g_J(b))$, $T^{\pm}=\tau^{\pm}(a)+\tau^{\pm}(b)$, $t^{\pm}=\tau^{\pm}(a)-\tau^{\pm}(b)$, and $T_z=\tau_z(a)+\tau_z(b)$.

Thus, the problem of calculating $\hat S^{--}_J$~(\ref{S}) is reduced to the problem of calculating the two-body double commutator $S^{2\nu}_J(ab)$~(\ref{S_J(ab)}). Bearing in mind a further calculation of $S^{2\nu}_J$ as a matrix element of $\hat S^{--}_J$ between antisymmetric nuclear wave functions of the initial and final nuclei, the operators of different permutation symmetry $g_J^\pm$ and $T^{\pm}$, $t^{\pm}$ acting in the two-body space are introduced.

By making use of isospin projection operators, the original two-body interaction, which is considered exactly isospin symmetric here, can be partitioned into two components, corresponding to different projections $T_z$ of the total isospin of the two-nucleon system:
\bea
&& v_{ab} = v_{ab}(T_z=0) + v_{ab}(|T_z|=1) \\
&& v_{ab}(T_z=0) \equiv v_{ab} \frac{(1-\tau_z(a) \tau_z(b))}{2}\ , \ 
v_{ab}(|T_z|=1) \equiv v_{ab} \frac{(1+\tau_z(a) \tau_z(b))}{2},
\eea
each of which can further be represented in terms of the interaction components corresponding to a definite total isospin $T=0,1$:
$ v_{ab}(T_z=0) = v_{ab}(T=1,T_z=0)\Pi_{T=1} + v_{ab}(T=0,T_z=0)\Pi_{T=0}$; $v_{ab}(|T_z|=1)= v_{ab}(T=1,|T_z|=1)$ ($\Pi_{T}$ is a corresponding total isospin projection operator).

The standard way of renormalization of the residual p-p interaction in the proton-neutron QRPA is:
\be 
v_{ab} \to g_{pp} v_{ab}(T_z=0) + g_{pair} v_{ab}(|T_z|=1),
\label{Vab_ren}
\ee
because only $v_{ab}(|T_z|=1)$ enters the BCS gap equations, while $v_{ab}(T_z=0)$ is responsible for the mixing of proton-neutron excitations (here for the sake of simplicity we consider the same $g_{pair}$ for proton and neutron subsystems).

For the F transitions ($g_{0}(a)=1$) one arrives at the following expression (see Appendix):
\be
S_0(ab)=2(g_{pair}-g_{pp})\left(T^{-}\right)^2 v_{ab}(T=1,T_z=0).
\label{S_F(ab)}
\ee
This result shows that the renormalization (\ref{Vab_ren}) obviously breaks the original isospin symmetry of the residual interaction if $g_{pp}\neq g_{pair}$ (that is usually the case in most of the realistic QRPA calculations; we shall later how this drawback can easily be remedied by a different renormalization that is more isospin consistent than the one of Eq.~(\ref{Vab_ren})).

The corresponding expression for the GT transitions ($g_{1}=\sss$) is more involved  (see Appendix) and reads
\bea
\label{S_GT(ab)}
S^{2\nu}_{1}(ab)&=&S^{2\nu}_{1}(ab,S=0)+ S^{2\nu}_{1}(ab,S=1)\\
S^{2\nu}_{1}(ab,S=0)&=&6\left(T^{-}\right)^2\left[g_{pair}v_{ab}(T=1,S=0)-g_{pp}v_{ab}(T=0,S=1)\right]\Pi_{S=0}\label{S'_GT(ab)}\\ 
S^{2\nu}_{1}(ab,S=1)&=&2\left(T^{-}\right)^2\left[(g_{pp}-g_{pair})v_{ab}(T=1,S=1)\right.\nonumber\\
&&\left.+g_{pp}(v_{ab}(T=1,S=1)-v_{ab}(T=0,S=0))\right](1-\Pi_{S=0}), 
\label{S_GT(ab)1}
\eea
where the operator $\Pi_{S=0}\equiv|00\rangle \langle 00|_{\ \hskip-0.25cm \phantom{|}_S}$ projects onto the spin $S=0$ state of two nucleons. The operators $v_{ab}(T,S)\equiv \langle TS|v_{ab}|TS \rangle$ are the expectation values of the original two-body interaction in two-body spin-isospin states $|TS \rangle$, and therefore only depend on the spatial coordinates of two nucleons. 

Now we proceed with calculations of $S^{2\nu}_{J}$ as the matrix elements of the two-body operators $\hat S^{--}_{J}$ between the ground states of the initial and final nuclei.
Since $v_{ab}$ is of a short range, 
then the nucleon pairs in the relative spatial $s$ wave must predominantly contribute to the $S^{2\nu}_{J}$. The operator $\left(T^{-}\right)^2$ in Eqs.~(\ref{S_F(ab)} and \ref{S_GT(ab)}) transforms a $T=1$ neutron pair into a $T=1$ proton pair [and becomes just a number $\left(T^{-}\right)^2=2$ for these states]. Such pairs of nucleons must then be in the state with the total spin $S=0$ to assure antisymmetry of the total two-body wave function. This means that the term $S^{2\nu}_{1}(ab,S=1)$ (\ref{S_GT(ab)1}) as projecting onto $S=1$ states can safely be neglected.
Further, one can anticipate in advance that the dominating contribution to $S^{2\nu}_{J}$ should come from the paired neutrons and protons in the two-body state $J^\pi=0^+$. 
By taking into account only this leading contribution of paired nucleons, one arrives at the following representation for $S^{2\nu}_{0}$:

\bea
S^{2\nu}_{0} &  = & (1 - g'_{pp}) g_{pair} S_0^{(pair)}, 
\label{S2nuF}\\
S_0^{(pair)} &  = & \sum_{pn} G(ppnn;J=0,T=1)
\langle 0^+_f|[c^\dagger_{p}c^\dagger_{p}]_{00}[c_{n}c_{n}]_{00}| 0^+_i\rangle,
\eea
where $g'_{pp}\equiv g_{pp}/ g_{pair}$, and $[c^\dagger_{t}c^\dagger_{t}]_{00}$ and $[c_{n}c_{n}]_{00}$ are bifermionic operators made of the coupled to $J^\pi=0^+$ particle creation and annihilation operators $c^\dagger_{t}, c_{t}$ ($t=p,n$), and the $G$ matrix $G(J=0,T=1)$ corresponds to the two-body interaction $v_{ab}(T=1)$ (from which only the component $v_{ab}(T=1,S=0)$ is active in the $0^+$ channel). 

The corresponding expression for $S^{2\nu}_{1}$ reads
\bea
& S^{2\nu}_{1}   =  (1 - \gamma_{_{1}} g'_{pp}) g_{pair} S_{1}^{(pair)}, 
\label{S2nuGT}\\
& S_{1}^{(pair)}   =  3 S_0^{(pair)},\ \ \ \ \ \ \gamma_{_{1}}\equiv S_{1}^{(pp)}/S_{1}^{(pair)},
\label{S2nuGT1}\\
& S_{1}^{(pp)}  =  \sum_{pn} G'(ppnn;J=0,T=1)
\langle 0^+_f|[c^\dagger_{p}c^\dagger_{p}]_{00}[c_{n}c_{n}]_{00}| 0^+_i\rangle,\label{S2nuGT2}
\eea
where the $G$ matrix $G'(J=0,T=1)$ corresponds to the two-body interaction $v_{ab}(T=0,S=1)\Pi_{S=0}\Pi_{T=1}$ in Eq.(\ref{S'_GT(ab)}).~\footnote{In such a deuteron-type channel with $T=0,S=1$, because of the tensor interaction, the attraction must be stronger than in the one with $T=1,S=0$, thus one can expect $\gamma_{_{1}}>1$ in Eq.(\ref{S2nuGT1}) prior to any calculation.}

Let us note that retaining only the contribution of $J=0$ paired nucleons in the g.s. wave functions corresponds to the calculation of the double commutator (\ref{S}), defining $S^{2\nu}$, in the QBA. In the QRPA this means that the commutator of two bifermionic operators is substituted by its expectation value in the BCS state, which is a $c$-number (cf. a derivation of the QRPA matrices $A$ and $B$, e.g., Ref.~\cite{RingSchuck80}). In fact, there are additional, beyond the QRPA, contributions to $S^{2\nu}$ from the pairs with $J>0$ in the correlated g.s., but they must be suppressed as the following arguments suggest. For the pairs with $J>0$, which are in the relative $s$ wave and may therefore essentially contribute to the sum rule, the total $J$ must then coincide with the total orbital momentum of the pair. 
However, as previous calculations of the $0\nu\beta\beta$-decay transition densities have shown~\cite{anatomy,anatomy1}, the contributions from $J>0$ pairs come from larger internucleon distances, substantially exceeding the short range of the $NN$-potential,  then that of $J=0$ (the latter peaks essentially at 1-2 fm). This in combination with the short range of $v_{ab}$ causes a suppression of the $J>0$ contributions. 

An estimate of such a suppression is beyond the scope of this paper and deserves a separate study. These contributions can be estimated if in a calculation of $0\nu\beta\beta$-decay one substitutes the Coulomb-like $r$-dependence of the neutrino potential by 
a corresponding $r$-dependence of the $T=1$ component of a $NN$ potential. 
Note, that even in the case of the long-range neutrino potential the $0^+$-pair contribution is by far the largest one.

For the BCS description of pairing, by taking the same BCS solution for the initial and final nuclei, one gets $\langle 0^+_f|[c^\dagger_{t}c^\dagger_{t}]_{00}| 0^+_i\rangle=\langle 0^+_f|[c_{t}c_{t}]_{00}| 0^+_i\rangle=\hat j_{t}u_{t}v_{t}$, where $u,v$ are the Bogoliubov coefficients, $u_{t}v_{t}=\frac{\Delta_{t}}{2E_{t}}$, and the pairing gaps $\Delta_{t}$ satisfy the gap equation:
\be
\Delta_{t} = \frac{g_{pair 
}}{\hat j_{t}}\sum_{t'} G(ttt't';J=0,T=1)\hat j_{t'} u_{t'}v_{t'}.
\label{Delta}
\ee

It is noteworthy that the same expressions (\ref{S2nuF})-(\ref{S2nuGT2}) for $S^{2\nu}_{J}$ can be obtained by considering the double commutator (\ref{S}) in the QBA, like it was used in the derivation of Ref.~\cite{Rodin05}. 
A compacter expression can be obtained for the pairing sum rule $S_J^{(pair)}$ by taking into account the gap equation Eq.~(\ref{Delta}):
\bea
& g_{pair}\, S_J^{(pair)}=\displaystyle 
\frac{1}{4}\sum\limits_{pn} \langle p\|\beta_{J}\|n\rangle^2\,
\frac{\Delta_p\Delta_n(E_p+E_n)}{E_pE_n}
\eea

\section{Results and Analysis}

Within the QRPA approach of Refs.~\cite{Rod05,anatomy,src09} we have computed $M^{2\nu}_{J}$~(\ref{Mbb}), $S^{2\nu}_{J}$~(\ref{S2nu}), $M^{2\nu}_{J~\{cl\}}$~(\ref{Mbb_cl}) along with $M^{0\nu}(J)$~(\ref{M--}) and $S^{0\nu}(J)$~(\ref{M0nu1}) for the $\beta\beta$ decays $^{76}$Ge$\rightarrow ^{76}$Se, $^{100}$Mo$\rightarrow ^{100}$Ru, and $^{130}$Te$\rightarrow ^{130}$Xe ($J=0$ and $J=1$ for the F and GT transitions, respectively).
The parametrization of the Woods-Saxon mean field is adopted from the spherical calculations of Refs.~\cite{Rod05,anatomy,src09}. For each of the nuclei two sizes of the s.p. basis, the small one (s.b.) and the large one (l.b.) in the notation of Refs.~\cite{Rod05,anatomy,src09}, are used in the calculations. The small s.p. space consists of 9 levels (oscillator shells $N$=3,4) for $A=76$ and 13 levels (oscillator shells $N$=3,4 plus $f + h$ states from $N=5$) for $A=100,130$. The large basis contains 21 levels for $A=76,100$ (all states from shells $N=1-5$), and 23 levels for $A={130}$ ($N=1-5$ and $i$ orbits from $N=6$).

As in Refs.~\cite{Rod05,anatomy,src09}, the nuclear Brueckner $G$~matrix, a solution of the Bethe-Goldstone equation with different nucleon-nucleon potentials (Bonn-CD, Bonn-C, Argonne V18 and Nijmegen I), is used as a residual two-body interaction. The results obtained with different $G$~matrices look pretty similar to each other, and in all the figures below only the results obtained with the Bonn-CD $G$~matrix are represented. 

First, the BCS equations are solved to obtain the Bogoliubov coefficients $u$ and $v$, the pairing gaps $\Delta$, and the chemical potentials. To correctly reproduce the experimental odd-even nuclear mass differences for both protons and neutrons in initial and final nuclei, four slightly different renormalization factors $g_{pair}$ in Eq.~(\ref{Delta}) are needed. Here, we approximate the single parameter $g_{pair}$ of the preceding section by taking the average value of these four factors.

As in Refs.~\cite{Rod05,anatomy,src09}, we set the particle-hole renormalization factor $g_{ph} = 1$ in the QRPA equations. The calculated energy of the giant GT resonance, which is essentially independent of the size of the s.p. basis, is well reproduced with such a choice of $g_{ph}$~\cite{Rod05,anatomy,src09}. One must say that a particular choice of $g_{ph}$ in the QRPA has no effect on the sum rules $S^{2\nu}_{1}$ and $S^{2\nu}_{0}$ (\ref{S2nu}), and $S^{0\nu}(GT)$ and $S^{0\nu}(F)$~(\ref{S0nu}). This can be seen from the general analytic expressions (\ref{S2nuF},\ref{S2nuGT}) (determined exclusively by the p-p interaction) and is confirmed by the direct calculations.

\begin{figure}[t]

\

\centerline{\includegraphics[scale=0.5]{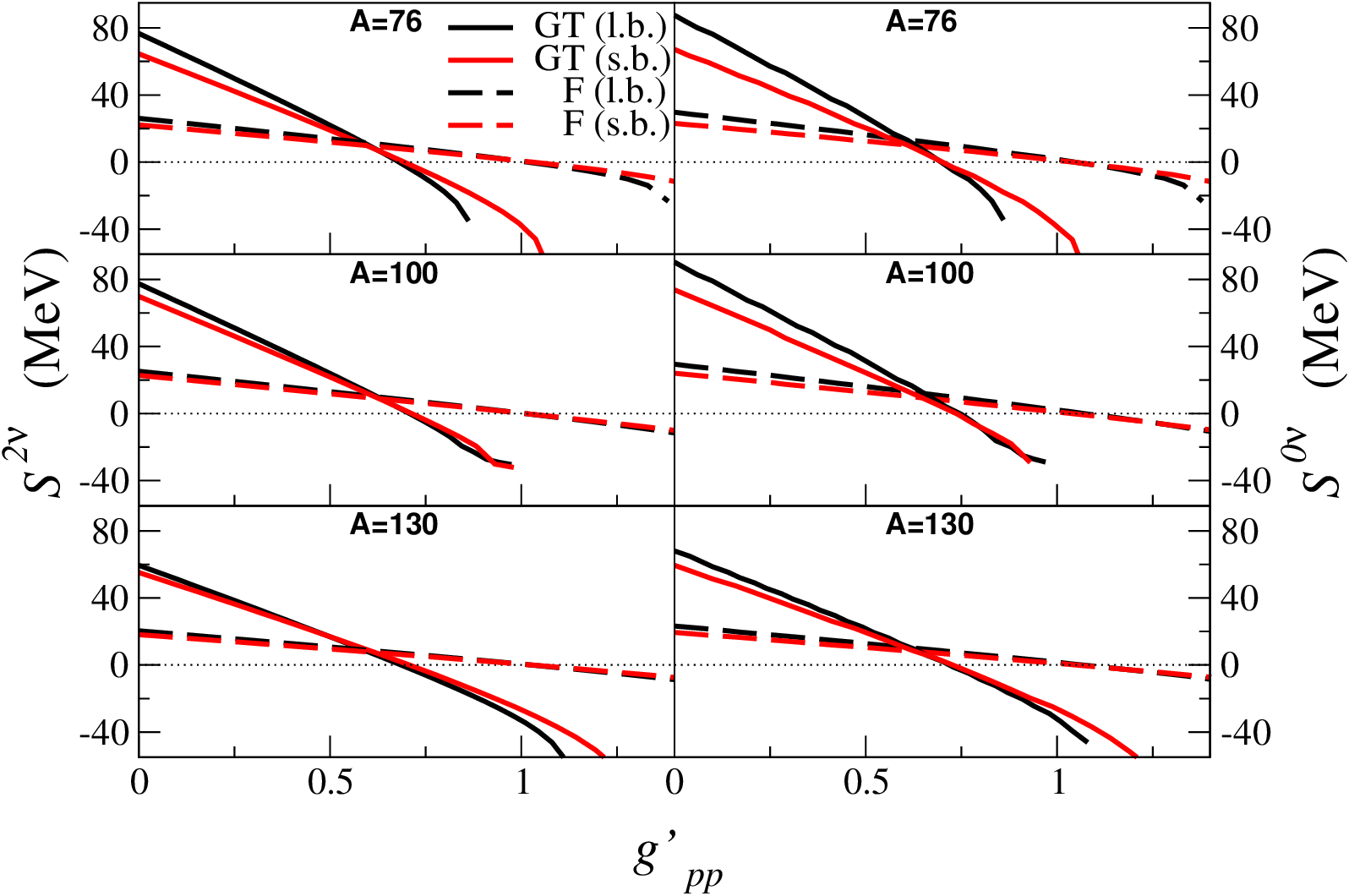}}
\caption{(Color online) The GT and F sum-rules $S^{2\nu}_{1}$ and $S^{2\nu}_{0}$ (\ref{S2nu}) (the solid and dashed lines in left column, respectively), and $S^{0\nu}(1)$ and $S^{0\nu}(0)$~(\ref{S0nu}) (the solid and dashed lines in right column, respectively) calculated within the QRPA approach of Ref.~\cite{Rod05,anatomy,src09}. The results for the small s.p. basis (s.b.) are represented by the red lines, and for the large one (l.b.) are represented by the black lines.} 
\label{fig.1}
\end{figure}

The sum rules $S^{2\nu}_{J}$ (\ref{S2nu}), and $S^{0\nu}(J)$~(\ref{S0nu}) calculated within the QRPA approach of Ref.~\cite{Rod05,anatomy,src09} are shown in Fig.~\ref{fig.1}.~\footnote{All the $0\nu$ quantities are calculated here by retaining only the leading Coulomb-like radial dependence in the neutrino potential} One can see a universal character of the almost perfectly linear dependencies $S(g'_{pp})$ for all the nuclei in question, with only a little dependence on the s.p. basis size. 
Only for $^{130}$Te$\rightarrow ^{130}$Xe does one observe a slight depletion of $S$ as a result of weaker proton pairing in $^{130}$Te (only 2 protons upon the $Z=50$ core). The fact that the dependence $S(g'_{pp})$ is almost the same for different nuclei can easily be understood from the phenomenological $A$ dependencies $\Delta\propto A^{-1/2}$ and $G_0\propto A^{-1}$ put into $S\propto \Delta_n\Delta_p/G_0$, Eq.~(\ref{S_separ}), for separable forces. These estimates also explain why $S(0)$ slightly increases with an enlargement of the s.p. basis [a smaller $g_{pair}$ (or $G_0$) is needed to fit the experimental $\Delta_n,\ \Delta_p$].

Also, $S_J^{2\nu}$ and the corresponding $S^{0\nu}(J)$ have a surprisingly good quantitative agreement (a possible clarification of this numerical observation needs a further study that is beyond the scope of this work). All the $S_F$ lines cross zero very close to $g'_{pp}=1$, in an excellent agreement with Eq.~(\ref{S2nuF}), despite the aforementioned numerical differences in the fitted $g_{pair}$ for protons and neutrons and the overlap factor used in the present QRPA calculation. The ratio $S_{1} = 3 S_F$ at $g'_{pp}=0$ comes out almost exact in the calculation, again in excellent accord with Eq.~(\ref{S2nuGT1}). All the $S_{1}$ lines cross zero at $g'_{pp}=\gamma^{-1}_{_{1}}<1$, 
as an expected result of a stronger attraction in the deutron-type channel $T=0,S=1$. The calculated values of $\gamma_{_{1}}$ are listed in Table~\ref{table.1}, here for different choices of the residual interaction.~\footnote{Here the same BCS vacuum of the initial nucleus is used also for the final one to avoid the influence of the overlap factor. The latter spoils the observed universality of $\gamma_{_{1}}$ in Table~\ref{table.1} by a few percent} For a given choice of the residual interaction, one sees again an impressive universality of this parameter, with the largest value $\gamma_{_{1}}\approx 1.5$ in the case of  the Bonn-CD $NN$-potential and the smallest value $\gamma_{_{1}}\approx 1.3$ in the case of the Bonn-C $NN$-potential.

\begin{table}[tbh]
\caption{The calculated parameter $\gamma_{_{1}}$ Eq.~(\ref{S2nuGT1}) for different nuclei, basis sizes and choices of the residual interaction.}
\label{table.1}
\begin{center}
   \begin{tabular*}{\linewidth}{@{\extracolsep{\fill}}lllll}   
\hline\noalign{\smallskip}
 & Bonn CD  &  Argonne V18 & Nijmegen I & Bonn C \\
 \noalign{\smallskip}\hline\noalign{\smallskip}
$^{76}$Ge (l.b.) & 1.485 & 1.418 & 1.371 &1.274 \\
$^{76}$Ge (s.b.) & 1.477 & 1.413 & 1.368 &1.280 \\
 \noalign{\smallskip}\hline\noalign{\smallskip}
$^{100}$Mo (l.b.) & 1.493 & 1.426 & 1.378 &1.279 \\
$^{100}$Mo (s.b.) & 1.498 & 1.426 & 1.378 &1.280 \\
 \noalign{\smallskip}\hline\noalign{\smallskip}
$^{130}$Te (l.b.) & 1.485 & 1.419 & 1.371 &1.276 \\
$^{130}$Te (s.b.) & 1.481 & 1.416 & 1.370 &1.280 \\ 
\noalign{\smallskip}\hline
\end{tabular*}
 \end{center}
\end{table}

In this connection, provided $S$ for GT transitions could be measured, it, as depending exclusively on the residual p-p interaction, would be a better quantity for fitting $g_{pp}$ than the experimental $M^{2\nu-exp}$. In fact, the absolute value of $S$ can be determined experimentally from charge-exchange reactions or single-$\beta$ decays if the single-state dominance for $2\nu\beta\beta$-decay is realized in one or another intermediate nuclear system. 

\begin{figure}[tbh]
\centerline{\includegraphics[scale=0.5]{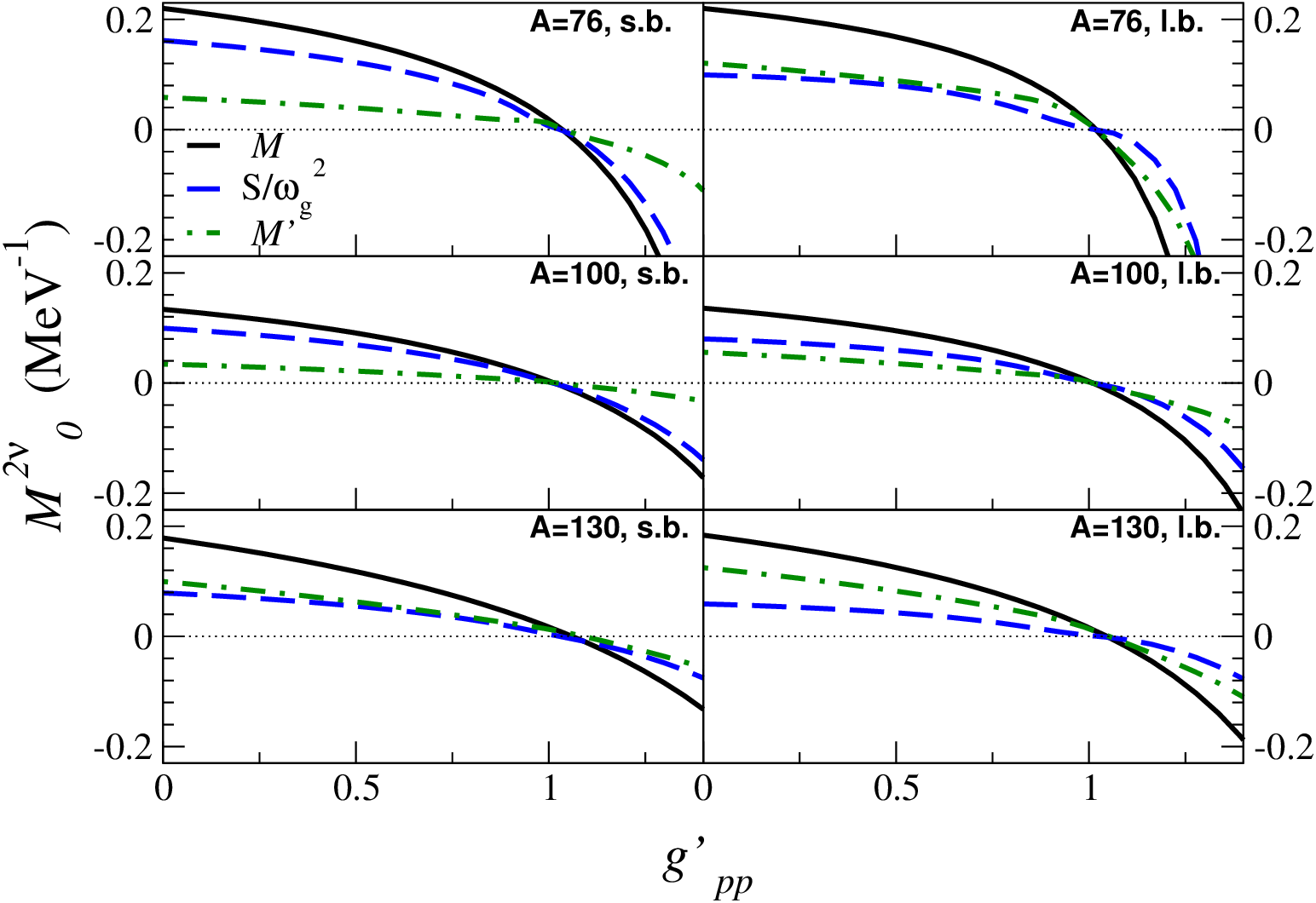}}

\

\

\

\centerline{\includegraphics[scale=0.5]{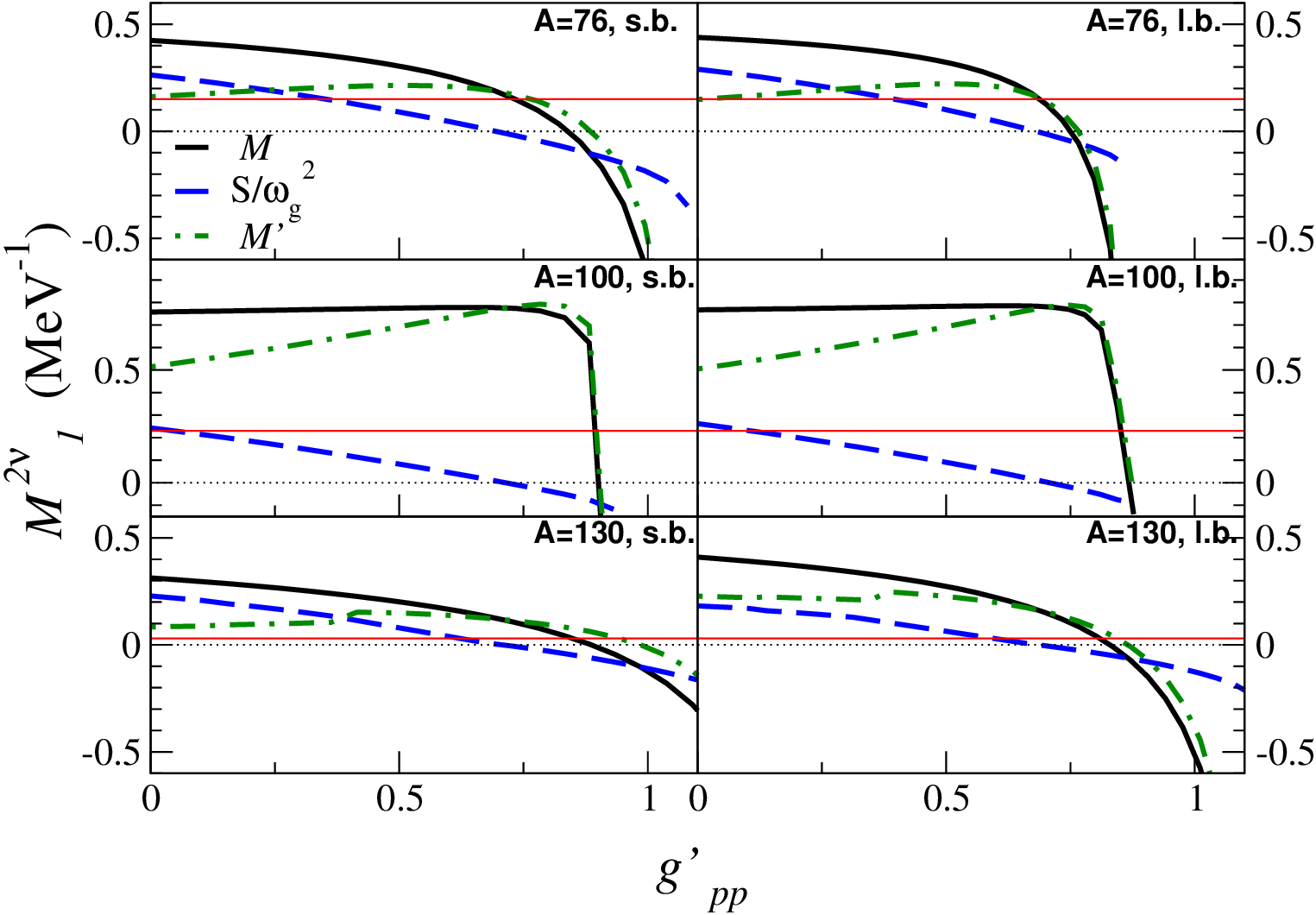}}
\caption{(Color online) $M^{2\nu}_{J}(g'_{pp})$ (solid line); $S^{2\nu}_J/\omega_g^2$ (dashed line); and their difference, ${M'}^{2\nu}_J(g'_{pp})$ (dot-dashed line), calculated within the QRPA according to Eqs.~(\ref{Mbb})-(\ref{S2nu}) with different basis sizes, s.b. and l.b.. The upper and lower panels contain the results for F and GT transitions, respectively. The thin solid horizontal lines in the lower panels represent the experimental values $M^{2\nu-exp}_{1}$ obtained in Ref.~\cite{Bar10} with the unquenched value of the axial-vector coupling constant $g_A=1.25$. } 
\label{fig.2}
\end{figure}

In Fig.~\ref{fig.2} we plot $M^{2\nu}_{J}(g'_{pp})$ (solid line), $S^{2\nu}_J/\omega_g^2$ (dashed line), and their difference, ${M'}^{2\nu}_J(g'_{pp})$ (dot-dashed line), calculated within the QRPA according to Eqs.~(\ref{Mbb})-(\ref{S2nu}). The upper and lower panels contain the results for F and GT transitions, respectively.
The value of $\omega_g$ is calculated as the mean energy of the GT or F strength distribution in the $\beta^-$ channel with a low-energy cut-off of 10 MeV.
One observes that  the dependencies $M^{2\nu}_{J}(g'_{pp})$ for different basis sizes, s.b. and l.b., plotted as functions of $g'_{pp}$ look much more similar to each other than
in the usual case when they are represented as functions of $g_{pp}$.
As for the function $S/\omega_g^2$, it shows basically a linear dependence on $g'_{pp}$ governed by the corresponding behavior of $S$.
One can see from the figure that the dependence ${M'}^{2\nu}_J(g'_{pp})$ is much smoother than the original one, $M^{2\nu}_{J}(g'_{pp})$, for realistic values of $g'_{pp}$ (apart from one exception --- $M^{2\nu}_{1}(g'_{pp})$ for $^{100}$Mo$\rightarrow ^{100}$Ru, where one approaches very close to the point of the QRPA collapse).
Also, all $M^{2\nu}_{0}$ cross zero very close to the point $g'_{pp}=1$ of restoration of the isospin symmetry of the renormalized residual interaction~(\ref{Vab_ren}).

\begin{figure}[t]
\centerline{\includegraphics[scale=0.5]{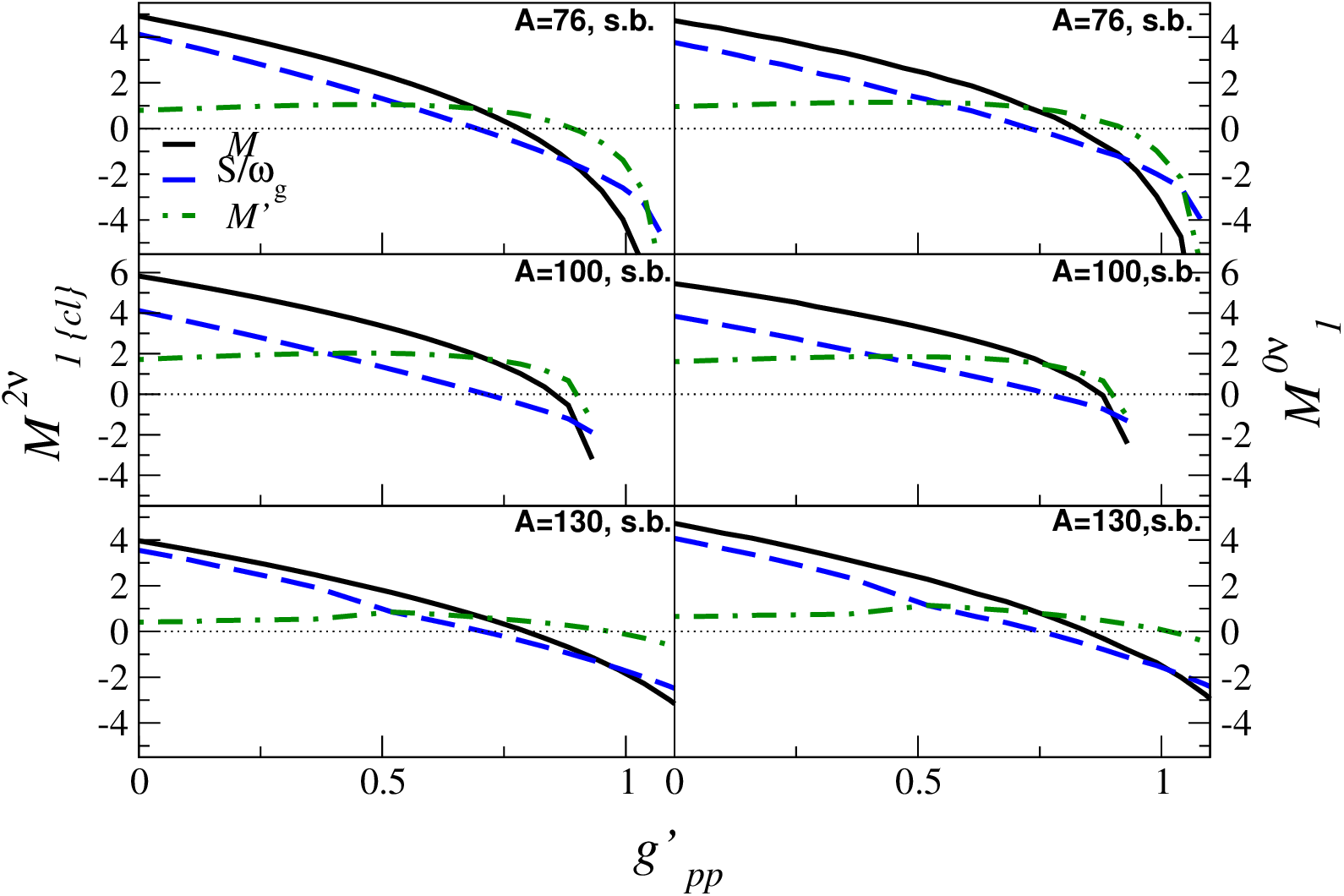}}

\

\

\

\centerline{\includegraphics[scale=0.5]{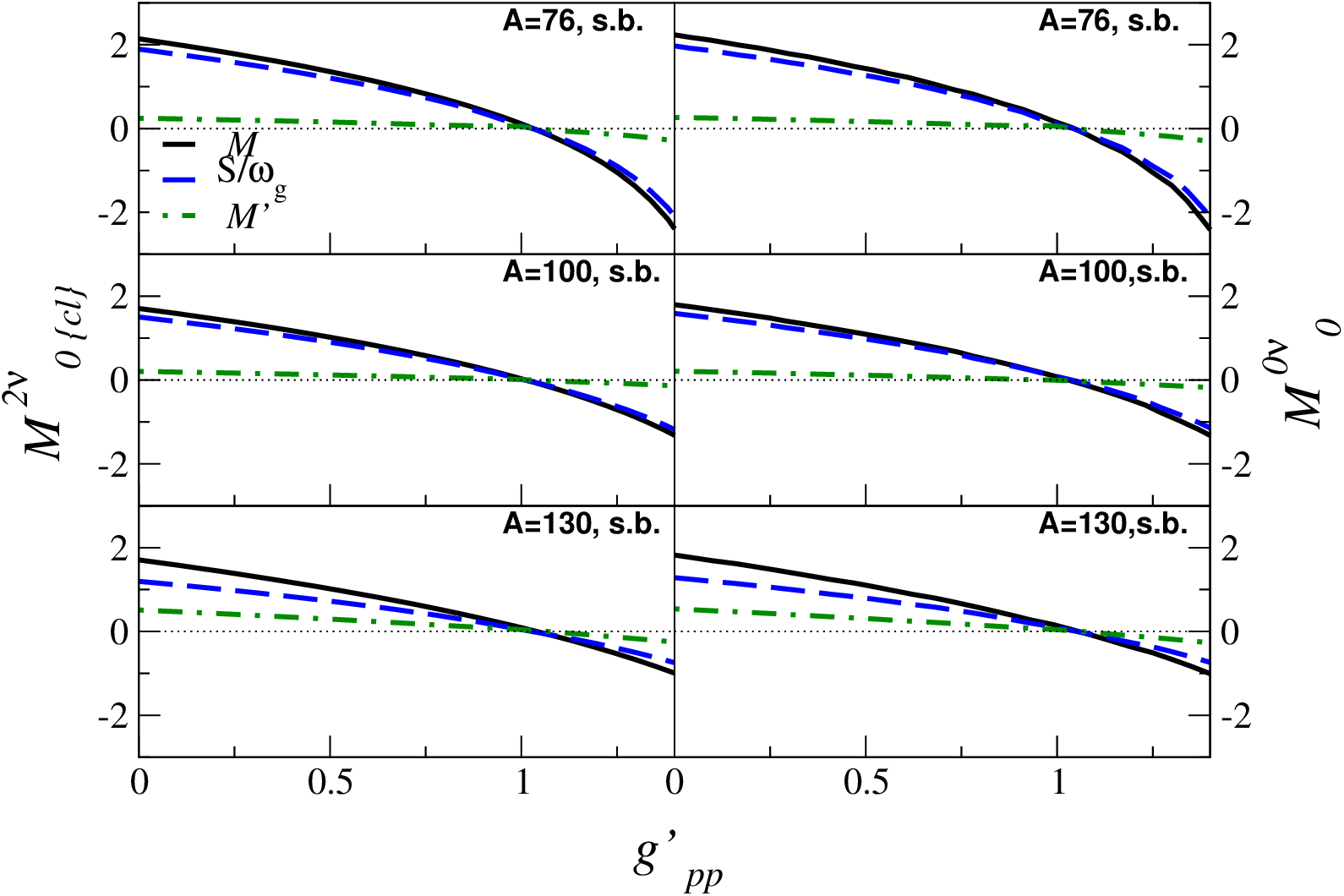}}
\caption{(Color online) The matrix elements $M^{2\nu}_{J~\{cl\}}$, $M^{0\nu}(J)$ (solid lines), $S^{2\nu}_J/\omega_g$, $S^{0\nu}(J)/\omega_g$ (dashed lines) and the difference, ${M'}^{2\nu~\{cl\}}_J$, ${M'}^{0\nu}(J)$ (dot-dashed lines), calculated within the QRPA according to Eqs.~(\ref{Mbb_cl})-(\ref{Mbb'_cl}) ($2\nu$) and Eqs.~(\ref{M0nu})-(\ref{M0nu1}) ($2\nu$) in the small basis (s.b.). The upper and lower panels contain the results for Fermi and Gamow-Teller transitions, respectively, and in each panel left and right columns of figures show the results for the $2\nu$ and $0\nu$ cases, respectively.
}
\label{fig.3}
\end{figure}

In Fig.~\ref{fig.3} we show for completeness the closure matrix elements $M^{2\nu}_{J~\{cl\}}$, $M^{0\nu}_{J}$ (solid lines), and $S^{2\nu}_J/\omega_g$, $S^{0\nu}_J/\omega_g$ (dashed lines) and the differences, ${M'}^{2\nu~\{cl\}}_J$, and ${M'}^{0\nu}_J$ (dot-dashed lines), calculated within the QRPA according to Eqs.~(\ref{Mbb_cl})-(\ref{Mbb'_cl}) ($2\nu\beta\beta$) and Eqs.~(\ref{M0nu})-(\ref{M0nu1}) ($0\nu\beta\beta$) in 
s.b.. The upper and lower panels again contain the results for F and GT transitions, respectively, and in each panel the left and right columns of figures show the results for the $2\nu\beta\beta$ and $0\nu\beta\beta$ cases, respectively. One sees that in this case almost the whole $g'_{pp}$ dependence of $M^{2\nu}_{J~\{cl\}}$ and $M^{0\nu}_{J}$ is governed by the $g'_{pp}$ dependence of $S^{2\nu}_J$ and $S^{0\nu}_J$, respectively.

The thin solid horizontal lines in the lower panels of Fig.~\ref{fig.2} represent the corresponding experimental values $M^{2\nu-exp}_{1}$ obtained in Ref.~\cite{Bar10} with the unquenched value of the axial-vector coupling constant $g_A=1.25$. 
Here one immediately sees a problem --- one needs $g'_{pp}<1$ to fit $M^{2\nu-exp}_{1}$, but for such a choice the renormalized residual interaction~(\ref{Vab_ren}) breaks isospin and $M^{2\nu}_{0}$ are spuriously large. However, this drawback is very easy to remedy by a slightly different prescription of a renormalization of the residual interaction. According to Eqs.~(\ref{S2nuF}, \ref{S2nuGT}), the F transitions are sensitive to $g_{pp}v_{ab}(T=1,S=0)$, 
whereas the GT transitions are sensitive to $g_{pp}v_{ab}(T=0,S=1)$. 
Therefore, it suffices to renormalize different $T$ components of $v_{ab}$ as
\be 
v_{ab} \to g_{pp} v_{ab}(T=0) + g_{pair} v_{ab}(T=1),
\label{Vab_ren1}
\ee
to have an isospin symmetric interaction, which allows one at the same time to fit the odd-even nuclear mass differences by means of $g_{pair}$ and $M^{2\nu-exp}_{1}$ by means of $g_{pp}$.
Direct QRPA calculations using the renormalization Eq.~(\ref{Vab_ren1}) show that the GT $g'_{pp}$ dependencies shown in Figs.~\ref{fig.1}--\ref{fig.3} stay practically the same (the  change would be barely visible in the figures), whereas the the F $g'_{pp}$ dependencies become constant, equal to the corresponding values at $g'_{pp}=1$ in Figs.~\ref{fig.1}--\ref{fig.3}. The effect of this new way of renormalization of the residual interaction on the total $0\nu\beta\beta$-decay NME $M^{0\nu}$ will be investigated elsewhere, but one can already anticipate that $M^{0\nu}_F$ will come out slightly smaller than in Ref.~\cite{Rod05,anatomy,src09}, whereas $M^{0\nu}_{GT}$ will barely be affected.

The results discussed above demonstrate very little dependence on the s.p. basis size. However, the smallest QRPA s.p. basis used in the analysis, $2 \hbar\omega$ for $A = 76$, is still much larger than the corresponding one of the SM which contains only four s.p. levels: $1p_{3/2}, 0f_{5/2}, 1p_{1/2}$ and $0g_{9/2}$~\cite{Poves}. The problem with this small $0\hbar\omega$ SM basis is that the spin-orbit partners $0f_{7/2}$ and $0g_{7/2}$ are missing, which leads to a strong violation of the Ikeda sum rule (ISR)~\cite{Ikeda} (the QRPA satisfies the ISR exactly, see a detailed discussion in Ref.~\cite{Esc10}). 

It is instructive to see what happens to the sum rules $S$ 
if one uses the SM s.p. basis. According to the arguments after Eq.~(\ref{S_GT(ab)1}), $S$
are mostly determined by the paired nucleons with $J=0$ in the g.s.. One can see in Fig.5 of Ref.~\cite{Esc10} that the contribution of $J=0$ pairs to $M^{0\nu}$ when calculated in the QRPA with the SM s.p. basis comes out in fairly good agreement with the corresponding result of a genuine SM calculation. This fact makes us confident that a  QRPA calculation of the sum rules $S$ with the SM s.p. basis should give a reasonably good estimate for the corresponding SM result.

The results of such a calculation are listed in Table~\ref{table.2}. It can be seen that the F sum rule $S_0$ is still reproduced well in the SM basis. However, the GT sum rules $S_{1}^{(pair)}$ and $S_{1}^{(pp)}$ come out strongly underestimated as a result of the missing  contributions to $S_{1}$ from the spin-flip transitions involving $0f_{7/2}$ and $0g_{7/2}$ s.p. states. The same happens to the parameter $\gamma_{_{1}}$, which is almost three times too small. 

Thus, one may state that the original inherent sensitivity of $S$ to the SU(4)-breaking part of the residual p-p interaction gets spuriously weak in the SM basis for $A=76$ system. To restore it, one must include the missing spin-orbit partners to the SM s.p. space, as, for instance, was the case in the SM description of $\beta\beta$ decay of $^{48}$Ca. 
The usual argument of the SM, that the weights of the admixtures of the missing states in the g.s. wave function are small, does not work here,
because the relatively small weights get compensated by large transition matrix elements to those states while calculating $S$.

\begin{table}[tbh]
\caption{Comparison of different $S$ calculated within the QRPA for $^{76}$Ge in the SM  basis (``4 levels") and s.b. (=``9 levels") (the Argonne-V18 $G$ matrix is used).}
\label{table.2}
\begin{center}
\begin{tabular*}{\linewidth}{@{\extracolsep{\fill}}lll}   
\hline\noalign{\smallskip}
   &  4 levels & 9 levels \\
 \noalign{\smallskip}\hline\noalign{\smallskip}
$S_0^{(pair)}$ & 13.8 & 15.8 \\
$S_{1}^{(pair)}$ & 21.9 & 46.9 \\
$S_{1}^{(pp)}$ & 12.3 & 66.3 \\
$\gamma_{_{1}}$  & 0.56 & 1.41\\
\noalign{\smallskip}\hline
\end{tabular*}
\end{center}
\end{table}

\begin{figure}[t]
\centerline{\includegraphics[scale=0.5]{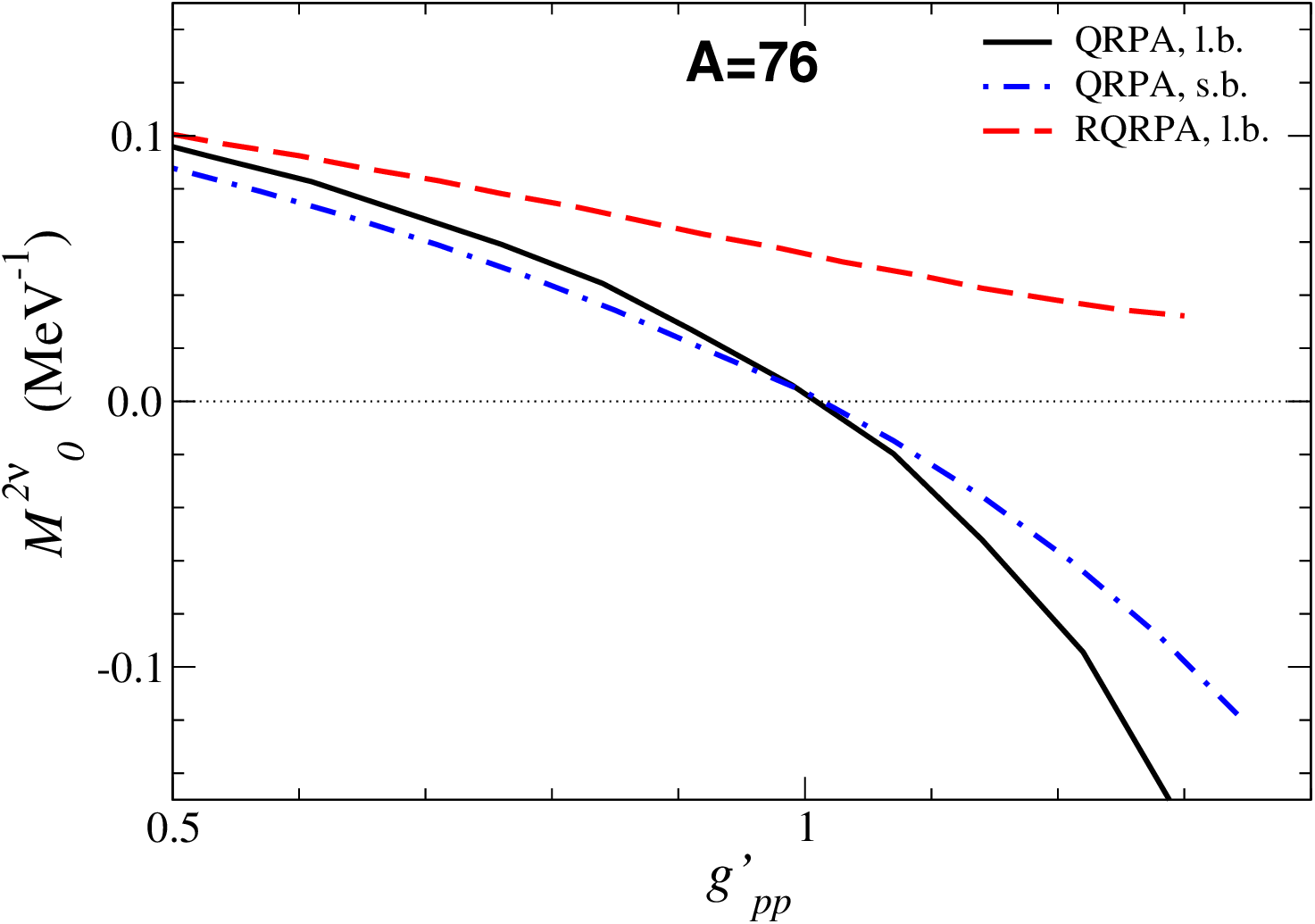}}
\caption{(Color online) $M^{2\nu}_{0}$ calculated within the RQRPA in comparison with the corresponding QRPA calculation.} 
\label{fig.5}
\end{figure}

To conclude the analysis of this section, a calculation of $M^{2\nu}_{0}$ within the renormalized QRPA (RQRPA)~\cite{Rod05,anatomy,src09} is performed and compared with the corresponding QRPA calculation  (Fig.~\ref{fig.5}). One can see that the calculated RQRPA dependence 
$M^{2\nu}_{0}(g'_{pp})$ does not cross zero at the physical point $g'_{pp}=1$ of the restoration of the isospin symmetry as the QRPA results do. Thus, this adds another  drawback of the RQRPA to the well-known violation of the ISR.

\section{Conclusion}
In the present work the sensitivity of the calculated $\beta\beta$-decay amplitudes to a realistic residual interaction is analyzed in the framework of the approach of Refs.~\cite{Rum98,Rodin05}. Both the GT and F matrix elements $M^{2\nu}$ for \bb decay, along with the monopole transition contributions to the total matrix elements $M^{0\nu}$ of \bbO decay, are calculated within the QRPA. 
Decompositions of $M^{2\nu}$ and $M^{0\nu}$ are obtained by the method of Refs.~\cite{Rum98,Rodin05} in terms of the corresponding energy-weighted sum rules $S$. It is shown that in most of the cases almost the whole dependence of $M^{2\nu}$ and $M^{0\nu}$ on the renormalization parameter $g_{pp}$ is accounted for by the $g_{pp}$ dependence of the corresponding sum rules $S$.
General expressions relating $S$ to a realistic residual p-p interaction are derived, which show a pronounced sensitivity of $S$ to the singlet-channel interaction in the case of F transitions and to the triplet-channel interaction in the case of GT transitions. In this connection, $S$ would provide the best quantity for fitting $g_{pp}$ if it could be measured (realistically, this can be done only in the case when the single-state dominance is realized in \bb decay).
Thus, the sensitivity of $M^{2\nu}$ and $M^{0\nu}$ to the SU(4)-symmetry-breaking part of the p-p residual interaction is dictated by the generic structure of the $\beta\beta$-decay amplitudes. Therefore, the choice of this part in a particular model and a further accurate calculation of its contribution to $S$ needs special attention.
Finally, a better isospin-consistent way of renormalization of the realistic residual particle-particle interaction to use in QRPA calculations is suggested.

\acknowledgments
The authors acknowledge support of the Deutsche Forschungsgemeinschaft within the SFB TR27 ``Neutrinos and Beyond''.

\appendix\section{Calculation of $S^{2\nu}_J(ab)$}

In the isospin space of the two-nucleon system $|TT_z\rangle$ ($T=0,1;T_z=0,\pm T$),  the isospin operators $T^{-}$ and $t^{-}$ can be expressed in terms of the isospin projection operators:
\bea
&&T^{-}=\sqrt{2}(|10\rangle \langle 11|_{\ \hskip-0.25cm \phantom{|}_T }
+|1-1\rangle \langle 10|_{\ \hskip-0.25cm \phantom{|}_T});~~~
t^{-}=\sqrt{2}(|00\rangle \langle 11|_{\ \hskip-0.25cm \phantom{|}_T}-|1-1\rangle \langle 00|_{\ \hskip-0.25cm \phantom{|}_T})
\eea

It is easy to verify that 
$T^{-} v_{ab} t^{-} = t^{-} v_{ab} T^{-} = 0$ 
as a consequence of the isospin conservation by strong interaction.
In addition, one has
$\tau_z(a) \tau_z(b) = \frac{T_z^{2}(ab)}{2}-1$, 
$T^{-}\tau_z(a) \tau_z(b)T^{-} = - \left(T^{-}
\right)^2$, and
$t^{-}\tau_z(a) \tau_z(b) t^{-} = \left(T^{-}
\right)^2$.
Then one finds that
$T^{-} v_{ab}(T_z=0) T^{-} = v_{ab}(T=1,T_z=0)\left(T^{-}\right)^2$,
$t^{-} v_{ab}(T_z=0) t^{-} = - v_{ab}(T=0,T_z=0)\left(T^{-}\right)^2$, and
$T^{-} v_{ab}(T_z=1) T^{-} = t^{-} v_{ab}(T_z=1) t^{-} = 0$,
and finally arrives at Eq.~(\ref{S_F(ab)}) for the F transitions.

For the GT transitions $g_{1}(a)=\sss_a$, and the operator 
$\tilde g_{1}(a)g_{1}(b)\equiv-\sss_a\sss_b$ 
has the eigenvalues $-2(S(S+1)-3)$ in the spin space of the two-nucleon system $|SM\rangle$, $S=0,1;M=0,\pm S$. The operators $g_{1}^+$ and $g_{1}^-$ are equal to 
$\vec S \equiv \frac12(\sss_a+\sss_b) $ and $\vec s \equiv \frac12(\sss_a - \sss_b)$, respectively , whose components can also be expressed in terms of spin projection operators:
\bea
&& S^{-}=\frac{1}{\sqrt{2}}(|10\rangle \langle 11|_{\ \hskip-0.25cm \phantom{|}_S}+|1-1\rangle \langle 10|_{\ \hskip-0.25cm \phantom{|}_S})\\
&& S^{+}=\frac{1}{\sqrt{2}}(|11\rangle \langle 10|_{\ \hskip-0.25cm \phantom{|}_S}+|10\rangle \langle 1-1|_{\ \hskip-0.25cm \phantom{|}_S})\\
&& S_z=\sum_s M|1M\rangle \langle 1M|_{\ \hskip-0.25cm \phantom{|}_S}
\eea
and
\bea
&& s^{-}=\frac{1}{\sqrt{2}}(|00\rangle \langle 11|_{\ \hskip-0.25cm \phantom{|}_S}-|1-1\rangle \langle 00|_{\ \hskip-0.25cm \phantom{|}_S})\\
&& s^{+}=\frac{1}{\sqrt{2}}(|11\rangle \langle 00|_{\ \hskip-0.25cm \phantom{|}_S}-|00\rangle \langle 1-1|_{\ \hskip-0.25cm \phantom{|}_S})\\
&& s_z=|10\rangle \langle 00|_{\ \hskip-0.25cm \phantom{|}_S}+|00\rangle \langle 10|_{\ \hskip-0.25cm \phantom{|}_S}.
\eea

Then one finds the following expressions:
\bea
&& 
\tilde g_{1}^- v_{ab}(T=0) g_{1}^-
= s_z v_{ab}(T=0) s_z + 2(s^{-} v_{ab}(T=0) s^{+} + s^{+} v_{ab}(T=0) s^{-})\\
&& = \left[\sum_s \langle 1M|v_{ab}(T=0)|1M\rangle_S \right] |00\rangle \langle 00|_{\ \hskip-0.25cm \phantom{|}_S} + 
\langle 00|v_{ab}(T=0)|00\rangle_S \left[ \sum_s |1M\rangle \langle 1M|_{\ \hskip-0.25cm \phantom{|}_S}\right]
\eea


and

\bea
&& 
\tilde g_{1}^+ v_{ab}(T=0) g_{1}^+
= S_z v_{ab}(T=1) S_z + 2(S^{-} v_{ab}(T=1) S^{+} + S^{+} v_{ab}(T=1) S^{-}) \\
&& = 2 \sum_s \langle 1M|v_{ab}(T=1)|1M\rangle_S |1M\rangle \langle 1M|_{\ \hskip-0.25cm \phantom{|}_S} 
= 2 v_{ab}(T=1,S=1) (1 - |00\rangle \langle 00|_{\ \hskip-0.25cm \phantom{|}_S}).
\eea
Finally, having collected all the contributions, one arrives at Eq.~(\ref{S_GT(ab)}) for the GT transitions.

\end{document}